\documentclass{epl2} 

\usepackage{amsmath,amssymb}

\newcommand{\D}{\Delta}
\renewcommand{\d}{\delta}
\renewcommand{\L}{\Lambda}
\renewcommand{\l}{\lambda}

\newcommand{\G}{{\mit\Gamma}}

\newcommand{\e}{\epsilon}

\newcommand{\s}{\sigma}
\renewcommand{\o}{\omega}
\renewcommand{\O}{\Omega}

\newcommand{\m}{\mu}
\newcommand{\n}{\nu}
\renewcommand{\r}{\rho}
\renewcommand{\s}{\sigma}

\newcommand{\p}{\pi}
\newcommand{\f}{\phi}

\renewcommand{\th}{\theta}
\renewcommand{\e}{\epsilon}

\newcommand{\erf}{\ensuremath{\mathop{\rm erf}\nolimits}}
\newcommand{\Tr}{\ensuremath{\mathop{\rm Tr}\nolimits}}
\renewcommand{\Re}{\ensuremath{\mathop{\rm Re}\nolimits}}
\renewcommand{\Im}{\ensuremath{\mathop{\rm Im}\nolimits}}
\newcommand{\realni}{\ensuremath{\mathbb{R}}}
\newcommand{\ds}{\displaystyle}

\title{Solution to the Cosmological Constant Problem in a Regge Quantum Gravity Model}

\author{A. Mikovi\'c\inst{1,2}\thanks{E-mail: \email{amikovic@ulusofona.pt}} \and M. Vojinovi\'c\inst{2}\thanks{E-mail: \email{vmarko@ipb.ac.rs}}}

\institute{                    
  \inst{1} Departamento de Matem\'atica, Universidade Lus\'ofona de Humanidades e Tecnologias - Av. do Campo Grande, 376, 1749-024 Lisboa, Portugal\\
  \inst{2} Grupo de Fisica Matem\'atica da Universidade de Lisboa - Av. Prof. Gama Pinto, 2, 1649-003 Lisboa, Portugal
}

\abstract{
We show that it is possible to solve the cosmological constant (CC) problem in a discrete quantum gravity theory based on Regge calculus by using the effective action approach and a special path-integral measure. The effective cosmological constant is given as a sum of 3 terms: the classical CC, the quantum gravity CC and the matter CC. Since the observations can only measure the sum of these 3 terms, we can choose the classical CC to be equal to the negative value of the matter CC. Hence the effective CC is given only by the quantum gravity CC, which is determined by the path-integral measure. Since the path-integral measure depends on a free parameter, this parameter can be chosen such that the effective CC gives the observed value.}

\begin{document}

\maketitle

\section{Introduction}

The cosmological constant problem, for a review and references see \cite{cc}, is the problem of explaining the presently observed value of the cosmological constant (CC) within a quantum theory of matter and gravitation. In any quantum gravity (QG) theory there should be a natural length scale, which is the Planck length $l_P \approx 10^{-35}\un{m}$. Consequently, the quantum correction to the classical value of CC should be of order $l_P^{-2}$. However, this natural theoretical value is $10^{122}$ times larger from the observed value, see \cite{cc}, and the problem is to explain this huge discrepancy. It is expected that an explanation should be provided by a well-defined QG theory. String theory has an explanation based on the landscape of string vacuua \cite{str}, but many physicists find this explanation unsatisfactory because it is a multiverse argument. Other known QG theories, like loop quantum gravity and spin-foam (SF) models, see \cite{lqg} and \cite{sfm} for reviews and references, as well as the casual dynamical triangulations \cite{cdt}, have not been able to provide an explanation.

Recently a generalization of SF models of QG was proposed, under the name of spin-cube (SC) models \cite{mv2p,scube}. The SC models were proposed in order to solve the two key problems of SF models: obtaining the correct classical limit and enabling the coupling of fermionic matter. This is achieved by introducing the edge lengths for a given triangulation of spacetime as independent variables and a constraint which relates the spins for the triangles with the corresponding triangle areas. A spin-cube model is equivalent to a Regge state-sum model (RSS), and it has general relativity (GR) as its classical limit \cite{scube}. A systematic study of the semiclassical approximation for RSS models was started in \cite{amr}, by using the effective action approach. It was also shown in \cite{amr} that an appropriate choice of the simplex weights, or equivalently by choosing the path-integral (PI) measure, one can obtain a naturally small CC, of the same order of magnitude as the observed value. However, the calculation in \cite{amr} did not take into account the contribution from the matter sector, and as is well known, the perturbative matter contributions to CC are huge compared to the observed value, see \cite{cc}.

\section{Effective action for matter and gravity}

In order to see what is the effect of matter on the value of CC we will consider a scalar field $\f$ on a 4-manifold $M$ with a metric $g$ such that the scalar-field action is given by
\begin{equation}
S_s (g,\f) = \frac{1}{2}\int_M d^4 x \sqrt{|g|}\left[g^{\m\n}\,\partial_\m \f \,\partial_\n \f -  U(\f) \right]\,, \label{sca}
\end{equation}
where $U(\f)$ is a polynomial of the degree greater or equal than 2.

When the metric $g$ is non-dynamical, the equations of motion of (\ref{sca}) are invariant under the constant shifts of the potential $U$. However, we know that the metric is dynamical, so that the constant shifts in $U$ will give contributions to the cosmological constant term. These classical shifts of the potential will affect the value of the classical cosmological constant $\L_c$, so that we will assume that $\L_c \ne 0$.

The QG theory we are going to use will be based on the assumption that the structure of spacetime at short distances is given by a piecewise linear manifold $T(M)$, which corresponds to a triangulation of $M$.  The classical geometry of $T(M)$ is described by a choice of the edge lengths $L_\e$, $1\le \e \le E$, which are positive and satisfy the triangle inequalities. The action (\ref{sca}) becomes
\begin{equation}
S_{Rs} = \frac{1}{2}\sum_\s V_\s (L)  \sum_{k,l} g^{kl}_\s (L)\,  \f'_k \, \f'_l - \frac{1}{2}\sum_\nu V_\nu^* (L)\, U( \f_\nu) \,,
\end{equation}
where $V_\s$ is the 4-volume of a 4-simplex $\s\in T(M)$, $g^{kl}_\s$ is the inverse matrix of the metric in $\s$ given by
\begin{equation}
g_{kl}^{(\s)} = \frac{ L_{0k}^2 + L_{0l}^2 -L_{kl}^2}{L_{0k}L_{0l}} \,,
\end{equation}
$ \f'_k = (\f_{\nu_k} - \f_{\nu_0})/L_{0k}$ and $V^*_\nu$ is the volume of the dual cell for a vertex point $\nu$ of $T(M)$, see \cite{rc}.

The quantum corrections due to gravity and matter fluctuations can be described by using the effective action, which will be based on the following classical action
\begin{equation}
S(L,\f) = \frac{S_{Rc}(L)}{G_N}   + S_{Rs} (L,\f) \,,\label{cla}
\end{equation}
where 
\begin{equation}
S_{Rc} = - \sum_{\D=1}^F  A_\D (L) \theta_\D (L) + \L_c  V_4 (L)
\end{equation} 
is the Regge action with a classical CC and $G_N$ is the Newton's constant. $A_\D (L)$ is the area of a triangle $\D\in T(M)$, $\th_\D$ is the deficit angle and $V_4$ is the 4-volume of $T(M)$. We will introduce a classical CC length $L_c$ given by $\L_c = \pm 1/L_c^2$.

The path integral for the action (\ref{cla}) is given by
\begin{equation}
Z = \int_{ D_{E}} \, \mu (L) \, d^E L \,\int_{{\realni}^V} \prod_{\nu=1}^V d\f_\nu \, e^{ i S(L,\f)/\hbar} \,, \label{crss}
\end{equation}
where the integration region $D_E$ is a subset of ${\realni}_+^E$ where the triangle inequalities hold. The measure $\m$ has to be chosen such that it makes $Z$ finite and $\m$ has to allow a semiclassical expansion for the effective action for large $L_\e$. If we also require to have the diffeomorphism invariance of the leading terms in the effective action when $E \gg 1$, then the simplest choice is
\begin{equation}
\m(L) = \exp (-V_4(L)/L_0^4 )\,, \label{pim}
\end{equation}
where $L_0$ is a free parameter, which will be determined by the observed value of the cosmological constant. $L_0$  has a dimension of a length, which is necessary in order to make $V_4 (L) /L_0^4$ dimensionless.

Since
\begin{equation}
\begin{array}{ccl}
S(L,\f)/\hbar &=& S_{Rc}(L)/l_P^2 + G_N S_{Rs} (L,\f)/l_P^2 \\
&=& S_{Rm}(L,\f)/l_P^2  \\
\end{array}
\end{equation}
the effective action (EA) equation becomes 
\begin{equation}
e^{\frac{i}{l_P^2}\G (L,\f)} = \int_{D_E (L)} d^E l \,\int_{{\realni}^V} d^V \chi
 \exp \left[ \frac{i}{l_P^2} \left( \bar S_{Rm} (L+l, \f + \chi)
 -\sum_\e \frac{\partial\G}{\partial L_\e }\,l_\e -\sum_\p \frac{\partial\G}{\partial \f_\p }\,\chi_\p \right) \right],\label{mpe}
\end{equation}
where we have introduced $\bar S_{Rm} = \bar S_{Rc} + G_N S_{Rs} (L,\f)$ and $\bar S_{Rc} = S_{Rc} +il_P^2 V_4/L_0^4$. The integration region $D_E (L)$ is a subset of ${\realni}^E$ obtained by translating the region $D_E$ by the vector $-L$, see \cite{amr}. The imaginary term in $\bar S_{Rm}$ comes from the measure (\ref{pim}). This measure also ensures that we can use the approximation $D_E(L) \approx {\realni}^E$ when $L_\e \to \infty$ in (\ref{mpe}) in order to solve it perturbatively in $l_P^2$, see \cite{amr}. The reason is that $L_\e$ are positive, so that when $L_\e\to\infty$
\begin{equation}
D_E (L) \approx [-L_1,\infty)\times\cdots \times [-L_E,\infty)\,.
\end{equation}
Consequently the Gaussian integral from QFT 
\begin{equation}
I = \int_{-\infty}^\infty e^{-zx^2/\hbar-wx} dx = \sqrt{\frac{\pi \hbar}{z}}\,e^{\hbar w^2 /4z}\,, 
\end{equation}
which generates the perturbative series in $\hbar$ because
\begin{equation}
\log I = C -\frac{1}{2}\log z + \frac{\hbar w^2}{4z}  \label{qfti}
\end{equation}
is analytic in $\hbar$, is replaced by the integral
\begin{equation}
I_L = \int_{-L}^\infty e^{-zx^2/\hbar -wx} dx \,.
\end{equation}
Since
\begin{equation}
I_L = \sqrt{\frac{\pi \hbar}{z}}\,e^{\hbar w^2 /4z}\left[\frac{1}{2} + \frac{1}{2} \,\erf \left( L\sqrt{\frac{z}{\hbar}}+ \frac{w}{2} \sqrt{\frac{\hbar}{z}}\right)\right] ,
\end{equation}
then
\begin{equation}
\log I_L = C  -\frac{1}{2}\log z + \frac{\hbar w^2}{4z} \vphantom{\int_M} + \sqrt{\frac{\hbar}{\pi z}} \frac{e^{-z\bar{L}^2/\hbar}}{2\bar L}\left[1 + O\left(\frac{\hbar}{z \bar{L}^2}\right) \right] ,\label{qgi}
\end{equation}
where $\bar L = L + \hbar w/2z$. Since (\ref{qgi}) is non-analytic in $\hbar$, this will introduce the non-perturbative terms of $O(1/\hbar)$ and we will not be able to solve the EA equation perturbatively. However, given that
$ \Re z = -(\log\m)''$, the nonperturbative terms will be suppressed for large $L_\e$ because $\Re (L,zL) \to +\infty$ for the measure (\ref{pim}). Therefore, the exponential path integral measure (\ref{pim}) will give a quantum theory with a well-defined semiclassical limit.

Given a classical action, the perturbative solution of the corresponding EA equation can be obtained by using the EA diagrams, see \cite{k}. It will be convenient to introduce a dimensionless field $ \sqrt{G_N}\,\f$, so that
$\sqrt{G_N}\,\f \to \f$ and $S_{Rm} = S_{Rc} + S_{Rs}$. The perturbative solution will be then given by
\begin{equation}
\G(L,\f) = S_{Rm} (L,\f) + l_P^2 \G_{1}(L,\f) + l_P^4 \G_{2}(L,\f) + \cdots\,, \label{pme}
\end{equation}
where $\G_{n}$ are given by the EA diagrams corrected by the measure contributions, see \cite{amr}.

We expect that the expansion (\ref{pme}) will be semiclassical for $L\gg l_P$ and $\f \ll 1$. This can be verified by studying the one-dimensional ($E=1$) toy model for the potential
$U(\f) = \frac{\o^2}{2} \f^2 + \frac{\l}{4!}\f^4$
where $\hbar\o = m$ is the matter field mass and $\l$ is the matter self-interaction coupling constant.
The toy-model classical action can be taken to be
\begin{equation}
S_{Rm} (L,\f) = \left(L^2 + \frac{L^4}{L_c^2}\right)\theta_r (L) + L^2 \left[\f^2 + \frac{L^2}{L_m^2} (\f^2 + a\f^4)\right]\theta_m(L) \,, 
\end{equation}
where $L_m = 1 /\o$, $\l/4! = a/L_m^2$, $\theta_r(L)$ and $\theta_m(L)$ are $C^\infty$ homogeneous functions of degree zero, while the PI measure can be taken to be $\m = \exp(-L^4 /L_0^4)$.

Note that the perturbative solution of an EA equation is a complex function, so that we need to perform a QG analog of the Wick rotation. This can be done by making a transformation $\G\to \Re\G \pm \Im \G$, see \cite{mvea}, so that the physical effective action will be given by 
\begin{equation}
S_{\rm eff} =(\Re\G \pm \Im \G)/G_N \,.\label{sef}
\end{equation}
The sign ambiguity will be fixed by requiring that the effective CC is positive.

\section{The effective cosmological constant}

The first-order quantum correction to the classical action (\ref{cla}) is determined by
\begin{equation}
\G_{1} =  i\frac{V_4}{L_0^4} + \frac{i}{2} \Tr \log \begin{pmatrix} 
 S_{LL} &  S_{L\f} \\ 
 S_{L\f} &  S_{\f\f}
\end{pmatrix} 
\,,
\end{equation}
where $S_{xy}$ are the submatrices of the Hessian matrix for $S_{Rm}$. Since 
\begin{equation}
\begin{array}{ccl}
S_{LL} & = & O(L^2)\,, \\
S_{L\f} & = & O(L^3)O(\f)\,, \\
S_{\f\f} & = & O(L^4)[1 +  O(\f^2)] \,, \\
\end{array}
\end{equation}
for $L$ large, then
\begin{equation}
\G_1 = i\frac{V_4(L)}{L_0^4} + \frac{i}{2} \Tr \log S_{LL} + \frac{i}{2} \Tr \log S_{\f\f} + O(\f^2) \,.\label{olm}
\end{equation}

The first term in (\ref{olm}) is the QG correction to the classical CC, while the matter sector will give a quantum correction to CC from the third term. This can be seen by considering the smooth manifold approximation when $E \gg 1$. In this case the third term in (\ref{olm}) can be calculated by using the continuum approximation
\begin{equation}
S_{Rs}(L,\f) \approx S_s (g,\f) \,,
\end{equation}
and the corresponding QFT in curved spacetime.  

Let us consider an edge-length configuration such that 
\begin{equation}
L_\e \ge L_{K} \gg l_P \,.\label{qftk}
\end{equation}
This condition ensures that the QG corrections are small and if $L_K \ll L_m$, we can calculate $\Tr \log S_{\f\f}$ by using the Feynman diagrams for $S_s$ with the UV momentum cutoff $\hbar /L_K = K$. Consequently the corresponding CC contribution will be given by the flat space vacuum energy density, since
\begin{equation}
\d\G_1 (L) \equiv \Tr \log S_{\f\f}\big{|}_{\f=0} \approx V_M \int_0^K dk \,k^3 \log (k^2 + \o^2 ) + \O_m(R,K) \,,
\end{equation}
where
\begin{equation}
\begin{array}{ccl}
\O_m (R,K) & = & \ds a_1 K^2 \int_M d^4 x \sqrt{|g|}\,R \\
& + & \ds \log\frac{K}{\o} \int_M d^4 x \sqrt{|g|}\left[ a_2 R^2 + a_3 R^{\m\n}R_{\m\n} +a_4 R^{\m\n\r\s} R_{\m\n\r\s}+ a_5\nabla^2 R \right] \vphantom{\int_M^M} \\
& + & \ds O\left(L_K^2 /L^{2}\right) \,, \vphantom{\int_M^M} \\
\end{array}
\end{equation}
see \cite{bd} for the values of the constants $a_k$. Therefore the only $O(L^4)$ term in $\d\G_1$ is
\begin{equation}
c_1 V_4 (L) \, K^4 \log\left(K/\o \right)  = c_1  \, \frac{V_4(L)}{L_{K}^4} \log(L_m /L_{K})\label{olt} \,,
\end{equation}
where $c_1$ is a numerical constant.

By using (\ref{sef}) and (\ref{olt}) we obtain that the one-loop CC is given by
\begin{equation}
\L_1 = \pm \frac{1}{2L_c^{2}} + \L_\m +  c_1 \, \frac{l_P^2}{2 L_{K}^4} \log(K/\o) \,, 
\end{equation}
where $c_1$ is a numerical constant of $O(1)$. We can write this as
\begin{equation}
\L_1 = \L_\m + \L_c + \L_m \label{ccfirstorder} \,.
\end{equation}

\section{Higher order contributions}

It is not difficult to see that the higher-order quantum corrections to CC will preserve the structure (\ref{ccfirstorder}), so that
\begin{equation}
\L = \L_\m + \L_c + \L_m \,.\label{lf}
\end{equation}
will be valid exactly.

The reason is that $\G (L,\f) = \G_g (L) + \G_m (L,\f)$ and
\begin{equation}
\G_m (L,\f) = V_4 (L)\, U_{\rm eff}(\f) \,,
\end{equation}
for constant $\f$, so that
the matter quantum fluctuations can only contribute to $\L$ additively. As far as the QG corrections are concerned, there are no corrections to $\L_{\m}$ beyond the one-loop order. This happens because the large-$L$ asymptotics of $\G_g (L)$ is determined by the asymptotics of
\begin{equation}
\log \bar S_{Rc}'' = \log O(L^2 /\bar L_c^2) + \log \theta_1 (L) + \log \left[ 1 + O(\bar L_c^2 /L^2)\right]  \,, 
\end{equation}
where $\th_1$ is a homogeneous function of degree zero and
\begin{equation}
\bar\G_{g,n} (L) = O\left(\left(\bar L_c^2 / L^4 \right)^{n-1}\right) \,,\label{dna}
\end{equation}
for $n>1$, where $\bar\G_{g,n}$ is the contribution from the $n$-loop EA diagrams for the action $\bar S_{Rc}$ and
\begin{equation}
\bar L_c^2 = L_c^2 \left( 1 + il_P^2 L_c^2 /L_{0}^4 \right)^{-1}\,.
\end{equation} 
Consequently
\begin{equation}
\G_{g,1}(L) = O(L^4/L_0^4) + \log O(L^2 /L_c^2) + \log \theta_1(L) + O(L_c^2 /L^2) \,, 
\end{equation}
and 
\begin{equation}
\G_{g,n} (L) = O((L_{0c}^2)^{1-n} L_c^2 /L^{2})\,,
\end{equation}
where $n>1$ and $L_{0c} = L_0^2 /L_c$. The perturbative solution is then valid for $l_P /L_{0c} < 1$, which is equivalent to $L_0 > \sqrt{l_P L_c}$. Consequently the QG corrections will be small if $L_\e \gg l_P$ and
\begin{equation}
L_0 \gg \sqrt{l_P L_c} \,.\label{scaa}
\end{equation}
The matter contributions will have a general form $\L_m = l_P^2 \, K^4 \, f(\bar\l , K /\o )$,
where $\bar\l = \l\, l_P^2$ and $f(x,y)$ is a $C^\infty$ function obtained by summing all one-particle irreducible vacuum Feynman diagrams for $U(\f)=\o\f^2/2 + \l\f^4/4!$ QFT.

The exact CC can be now written as
\begin{equation}
\L =\L_\m +\L_c + \L_m = \frac{l_P^2}{2L_0^4} \pm  \frac{1}{L_c^2} +  l_P^2 \, K^4 \, f(\bar\l , K /\o ) \,.
\end{equation}
The dependence of $\L$ on the free parameters $L_0$ and $L_c$ is such that we can set
\begin{equation}
\L_c + \L_m =\pm  \frac{1}{L_c^2} +  l_P^2 \, K^4 \, f(\bar\l , K /\o ) = 0 \,,\label{mcc}
\end{equation} 
so that $L_c$ is determined by $\L_m$. Then the condition (\ref{mcc}) gives
\begin{equation}
\L = \L_\m = \frac{l_P^2}{2L_0^4} \,.\label{qgl}
\end{equation}
Note that $\L > 0$ if we choose the plus sign in (\ref{sef}).

The equation (\ref{qgl}) will determine the parameter $L_0$, since the observed value of $\L$  is given by $l_P^2 \L \approx 10^{-122}$, so that $L_0 \approx 10^{-5}\,{\rm m}$.
Note that this value of $L_0$ satisfies $L_0 \gg l_P$, which is consistent with the condition (\ref{scaa}) for the validity of the semi-classical approximation. This agrees with the fact that CC can be observed only in the semiclassical regime of a QG theory.

Note that one can choose $\L_c + \L_m = C/l_P^2 $, where $C$ is an arbitrary number, so that $\L =\L_\m + C/l_P^2$. This gives $L_0/l_P = (\L l_P^2 - C)^{-1/4}$, and the only restriction on the value of $C$ is that $L_0(C) \gg l_P$. This restriction gives $-1 \ll C < 10^{-122}$ so that one can choose the natural value $C=0$. The value $C=0$ is considered natural because it can be a consequence of some principle or a symmetry.

Choosing $\L_c + \L_m = 0$ is not a fine tuning because we do not need to know the values of $\L_c$ and $\L_m$. In the usual QFT approach to the CC problem, it is assumed that there is no a quantum gravity contribution and $\L_c = 0$ so that $\L = \L_m$. Since $\L_m$ depends on the cutoff $K$, then at each order of the perturbation theory one has to adjust $K$ in order to obtain the observed value of $\L$. This is difficult to realize, because a natural cutoff $K = 1/l_P$ gives contributions of $O(1)$, while the desired value is of $O(10^{-122})$.

One could assume that $\L_c \ne 0$ and that QFT can produce a non-perturbative value for $\L_m$. Then $\L =\L_c + \L_m = C/l_P^2$ and one could take $C$ which gives the observed value of $\L$.
However, this approach is not satisfactory, because it gives $C=O(10^{-122})$ and does not explain why $\L_c l_P^2$ cancels $\L_m l_P^2$ to 122 decimal places. On the other hand, if $\L$ was zero, then $C=0$, and this value would not be a problem, because, as we mentioned earlier, $C=0$ could be a consequence of some symmetry or a principle, like supersymmetry, where $\L_c =\L_m =0$. However, in the real world $\L > 0$ and supersymmetry is broken, and consequently the approach based on a supersymmetric QFT has not produced a solution to the CC problem.

\section{Conclusions}

We have shown that the CC problem can be solved in a discrete QG theory based on the Regge formulation of GR. In this case the QG contributions to CC can be calculated explicitly, and they are given by a simple expression (\ref{qgl}). The matter contributions to CC are given by the sum of 1PI Feynman vacuum diagrams for the matter QFT with a physical momentum cut-off $\hbar /L_K$, where $L_K \gg l_P$. This contribution cannot be calculated explicitly, but it will have some definite value $\L_m$, because our theory is based on a finite path integral (\ref{crss}) and the effective action is defined non-perturbatively via the equation (\ref{mpe}). Since $\L$ is given by (\ref{lf}), we can choose $\L_c$ such that $\L_c = -\L_m$ so that $\L = \L_\m = l_P^2 /2L_0^4$, where $L_0$ is a free parameter from the path integral measure. By choosing $L_0 \approx 10^{-5}\,{\rm m}$ we obtain the presently observed value of the CC. This value of $L_0$ is natural for our approach, because it satisfies $L_0 \gg l_P$, so that it is consistent with the condition $L_0 \gg \sqrt{l_P L_c}$ for the validity of the semiclassical approximation. 

Note that in the standard approach to the CC problem, see \cite{cc}, the CC value is determined solely by the quantum fluctuations of matter and the classical value, so that $\L = \L_m + \L_c$. From a QG perspective, this is an oversimplification and the reason why the CC problem appeared. If the QG contribution is ignored, one then encounters the problem of how to arrange the cancellation of the matter contributions to 122 decimal places, by summing terms which are of $O(1)$, since the natural cut-off in the corresponding QFT is $L_K = l_P$. In our approach, we also use a QFT, but our QFT is an effective QFT, see \cite{wein}, since it is an approximation for a more fundamental theory. Hence we can take $L_K \gg l_P$ and therefore $l_P^2 \L_m = O(l_P^4 /L_K^4) \ll 1$. However, our $\L_m$ is still much bigger than the observed value, since $L_K < 10^{-20}\,{\rm m}$. This is because $L_K$ is a scale where QG corrections are small and the usual perturbative QFT is still valid, and from the LHC experiments we know that QFT is valid at $ 10^{-20}\,{\rm m}$. But when we take into account the QG contributions to CC and a non-zero classical CC, this problem is solved by canceling the matter contribution by appropriately choosing the value of the classical CC. Note that the individual terms in (\ref{lf}) cannot be observed and only the sum can be measured. For example, the Casimir experiment does not measure $\L_m$ but it measures the force $F \propto  U'_{\rm eff} (\f)$. In exact SUSY theories $\L_m = 0$, so that our mechanism can still work by choosing $\L_c = 0$, since (\ref{qgl}) is still valid for $\L_c =0$, see \cite{amr}.

Note that the transformation
\begin{equation}
\L_c + \L_m \to \L_c + \L_m + C'/l_P^2 \,,\qquad \L_\m \to \L_\m - C'/l_P^2 \label{shs}
\end{equation}
does not change $\L$. However, it is not clear whether (\ref{shs}) is a symmetry of the effective action. If (\ref{shs}) is a symmetry, we can choose any value of $C$, as long as $L_0 (C) \gg l_P$. However, it is more likely that (\ref{shs}) is not a symmetry, so that we will have a class of theories whose observables may depend on $C$. Since $C=0$ is an allowed value, it is the best candidate for a preferred value. In order to test how our theory depends on the value of $C$, one would have to find observables which depend on $C$ and compare them with the observed values. This can be done by computing more terms of the perturbative effective action.

\acknowledgments

We would like to thank M. Blagojevi\'c and V. Ra\-do\-va\-no\-vi\'c for discussions. This work has been partially supported by GFMUL and the FCT project ``Geometry and Mathematical Physics''. MV was supported by the FCT grant SFRH/BPD/46376/2008, and partially by the project ON171031 of the Ministry of Education, Science and Technological Development, Serbia.

\end{document}